\documentstyle[11pt,moriond]{article}

\def\be{\begin{equation}}
\def\ee{\end{equation}}
\def\bea{\begin{eqnarray}}
\def\eea{\end{eqnarray}}

\def\Msun{\hbox{$\rm\thinspace M_{\odot}$}}

\begin{document}
\vspace*{4cm}
\title{PROBING THE HOT IGM WITH AXAF}

\author{CLAUDE R. CANIZARES \& TAOTAO FANG }

\address{Department of Physics \& Center for Space Rsearch,\\Massachusetts
Institute of Technology,\\Cambridge MA USA 02136}

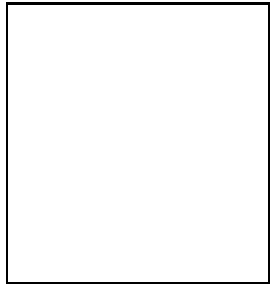
\maketitle\abstracts{The transmission grating spectrometers on AXAF
have sufficient sensitivity and energy resolution to detect resonance
X-ray absorption lines in the spectra of distant quasars.
Such lines would be produced in an ionized, and somewhat chemically
enriched  component of the inter-galactic
medium. The bulk of the baryonic content of the universe at redshifts
$\leq2-3$ could be in this form and would thus be revealed by AXAF.}

The Advanced X-ray Astrophysics Facility (AXAF)\footnote{
http://asc.harvard.edu}, now scheduled for launch
in late 1998, will carry two transmission grating
spectrometers
 that give
unprecedented energy resolution and good sensitivity over 0.07-8 keV.
These are the Low Energy\cite{pr}
 and High Energy Transmission
Gratings\cite{ma} (LETG and
HETG)\footnote{for HETG see http://space.mit.edu/HETG, for
LETG see http://www.rosat.mpe-garching.mpg.de/axaf},
 each of which is optimized for a portion of the energy band, and
each of which can achieve resolving powers of up to $E/\Delta E \approx1000$.

We plan to use AXAF's spectroscopic capability to probe the hot component
of the intergalactic medium (IGM).  This is an analog of the well
established studies of cooler IGM material through UV and optical
spectroscopy, which have revealed the Lyman $\alpha$ forest, damped Ly
$\alpha$ clouds, metal line systems, etc.  As in those studies, we will be
looking for spectral features imposed by the intervening material on the
X-ray continuum of distant quasars.

In the X-ray band, the features of interest are resonance lines from highly
ionized heavy elements, some of which are listed in Table 1.  Several of
these (marked with an asterix) are the Ly $\alpha$ lines of hydrogenic
species, whose energy is $13.6~Z^2$ eV (where Z is the atomic number).
Some other K lines of He-like alpha elements and L lines of Ne-like to
H-like Fe may also be of interest.  In principle, absorption edges from
bound-free transitions of heavy elements in any ionization stage should
also be present, but in practice the optical depths for edges are likely to
be too small.\cite{sh}

\begin{table}[t]
\caption{Resonant X-ray Absorption Lines  (* denotes Ly $\alpha$ lines of hydrogenic ions)}

\vspace{0.4cm}
\begin{center}
\begin{tabular}{|l c|l c|}
\hline
& & & \\
Ion & E (keV) & Ion & E (keV) \\
\hline
 OVII & 0.57 & MgXII* & 1.47 \\
 OVIII* & 0.65 & SiXIII & 1.87 \\
 FeXVII & 0.83 & SiXIV* & 2.01 \\
 NeIX& 0.92 & FeXXV & 6.70 \\
 NeX* & 1.02 & FeXXVI* & 6.97 \\
 MgXI & 1.35 & & \\
& & &  \\ 
\hline
\end{tabular}
\end{center}
\end{table}

The sensitivity of the AXAF spctrometers is such that we will readily
detect absorption lines with equivalent widths comparable to our spectral
resolution element of $\approx 1 eV$ (or $10^{-2} A$) at 1 keV.
One might push as much as an order of magnitude lower in selected
observations.  To produce such a feature requires a column density of the
appropriate ion of $\approx 10^{16} cm^{-2}$ which in turn implies hydrogen
column densities of $\approx 2\times10^{20} f_{ion}^{-1}[A/A(Fe)_{\odot}]^{-1}$,
where $f_{ion}$ is the ionization fraction and $A/A(Fe)_{\odot}$ is the
element abundance relative to the solar value for Fe.  To put this in an
astrophysical context, it would suffice to have $3\times10^{12}\Msun$ spread
over a sphere of radius 0.5 Mpc if it had 0.3 solar abundance; this would
correspond to a density $2.5\times10^{-4}$, or an overdensity of 50 times
critical at the present epoch.

Thus AXAF spectrometers will probe any moderately dense regions along the
line of sight containing material that is both ionized and somewhat
enriched with heavy elements.  Ionization could be either collisional or
radiative, corresponding to thermal plasma with tempratures above
$\approx3\times10^6K$ or ionization parameters $\xi\geq0.1$, respectively.

There is good reason to believe that ionized, chemically enriched regions
not only exist, but may well represent most of the baryonic content of the
universe at redshifts less than 2-3.\cite{gi}~\cite{fu}
Standard big bang nucleosynthesis
predicts baryon densities of $\Omega_B\approx 0.1-0.3$, and recent
estimates of the number and density of Ly $\alpha$ clouds at high redshift
are consistent with this value.  But $\Omega_B$ in stars or cool clouds at
lower redshift is less than 0.004  (values quoted are for $H_o=70 km~s^{-1}~Mpc^{-1}$).
Scaling from groups and clusters of galaxies supports the hypothesis that,
at lower redshift, most of the baryons are in the form of hot plasma.  It
is also plausible that this plasma is at least moderately enriched with
heavy elements.\cite{fu}~\cite{sa}~\cite{re}  Even at high z, the Ly $\alpha$ clouds appear to have metalicities
as high as $\approx 0.1$ solar.\cite{st}

It is instructive to consider the densest clouds seen in the optical/UV,
namely damped Ly $\alpha$ systems (DLA).  These have hydrogen column
densities approaching $10^{21} cm^{-2}s^{-1}$ and metalicities 10\% of
solar.
\cite{lu}  We estimate that the X-ray background, though highly uncertain, is
ten times too weak to photoionize a typical DLA sufficiently to give 
detectable X-ray lines.\cite{fa}  On the
other hand, it may well be that the clouds are also ionized
by shocks and/or by a local source of photoionizing radiation, as suggested by
the detection of OVI
lines.\cite{lu2} In that case the DLA could be seen in X-rays as well.

Observationally, rather than focus on the few DLA quasars already detected
in X-rays, we have chosen to take an approach that simply maximizes the
sensitivity to randomly distributed X-ray absorbers.  During the first
several months of the AXAF mission, we will observe the brighest X-ray
quasars with $z\geq2$, 0836+7104 and 2149-304.  Lower redshift abosrbers will be
probed by several observations of more local quasars and BL Lac objects.

Although our program is something of a fishing expedition, it has the
considerable potential of providing an entirely new probe of the IGM
and of revealing the bulk of the baryons in the universe.

\section*{Acknowledgments}
This work supported in part by NASA contract NAS8-38249

\end{document}